# Spectrally-engineered photonic molecules as optical sensors with enhanced sensitivity: a proposal and numerical analysis

**Svetlana V. Boriskina**

*School of Radiophysics, V. Karazin Kharkov National University, Kharkov 61077, Ukraine*
SBoriskina@gmail.com

We report a theoretical study of clusters of evanescently-coupled 2D whispering-gallery (WG) mode optical micro-cavities (termed "photonic molecules") as chemosensing and biosensing platforms. Photonic molecules (PMs) supporting modes with narrow linewidths, wide mode spacing, and greatly enhanced sensitivity to the changes in the dielectric constant of their environment and to the presence of individual sub-wavelength-sized nanoparticles in the PM evanescent-field region are numerically designed. This type of optical biosensor can be fabricated in a variety of material platforms and integrated on a single chip that makes it a promising candidate for a small and robust lab-on-a-chip device. Possible applications of the developed methodology and the designed PM structures to the near-field microscopy, single nano-emitter microcavity lasing, and cavity-controlled single-molecule fluorescence enhancement are also discussed.

OCIS codes: 130.3120, 230.5750, 140.4780, 130.6010;   PACS numbers:  42.60.Da, 42.79.Pw, 42.82.-m, 07.07.Df

## Introduction

The growing need for miniaturized instruments for medical, food and environmental testing together with raising level of bioterrorism concerns have fueled the development of optical biosensors. Optical microcavities (microdisks and microspheres) supporting high-Q WG modes (with light confined near the resonator surface by almost total internal reflection) have already been demonstrated to have great potential in the development of inexpensive, ultra-compact, highly sensitive and robust bio- and chemical sensors[1-9]. As compared to more traditional linear optical waveguide or fiber biosensors, microresonator-based devices benefit from much smaller size (several μm rather than a few centimetres) and higher sensitivity. Optical resonators enable significant decrease of analyte sample needed for efficient sensing.

Two basic sensing principles are usually employed in microresonator-based optical sensors. First class of sensors are designed to detect the resonant frequency shift caused by the change of the refractive index of the surrounding environment through the interaction of the evanescent field of the WG-mode outside the microcavity with the biochemical agent (mass sensing)[2-8]. This refractive index change can be caused by the change of the concentration of bio(chemical) material on the resonator surface or in the surrounding solution. Detection can also be made by measuring the output intensity change from the microresonator at a fixed wavelength. The other class of devices uses the evanescent field of the microresonator to excite the spectroscopic (fluorescence or Raman) signal from the molecules of the analyte (fluorescence sensing)[9,10]. In both methods, high Q-factors (narrow resonant linewidths and long cavity decay lifetimes) of the resonator modes are crucial for achieving high sensor sensitivity.

Microdisk resonators offer certain advantages over microspheres for use as biosensors as they can be fabricated with standard microfabrication technologies and easily integrated with other components and bus waveguides on a semiconductor chip. However, similar to the case of microspheres, there are several factors that may hinder the use of WG-mode circular microdisk resonators as sensitive and robust biosensing platforms. First of all, for a high-Q WG mode in a single microdisk, the interaction of the evanescent mode field with the analyte is weak due to high confinement of the modal field inside the resonator. Furthermore, the wavelength spacing between high-Q WG modes in optically-large cavities becomes so small that the individual modes cannot be resolved. They form a relatively broad peak in a measured cavity spectrum[8], making efficient detection difficult. On the other hand, widely spaced WG modes supported by wavelength-scale nano-cavities have low Q-factors, whose values decrease exponentially with decreased mode azimuthal order[11]. The symmetry of circular microdisks also results in double degeneracy of the WG modes they support (corresponding to the mode angular field dependence $\cos(m\varphi)$ or $\sin(m\varphi)$ ). This degeneracy is often removed by fabrication imperfections that break the symmetry of the structure and cause appearance of parasitic peaks in the resonator spectra[12].

To overcome the aforementioned limitations, we suggest using clusters of coupled nano-cavities (photonic molecules)[13] with wide free spectral range tuned to a symmetry-enhanced high-Q WG mode resonance[14] instead of a larger single microcavity. Properly designed photonic molecules (PMs) have already demonstrated high potential for the design of low-threshold microlasers[14-17]. In a previous publication[14], we have shown that by arranging circular WG-mode microcavities into pre-designed symmetrical configurations it is possible not only to preserve but also to significantly enhance their attractive features, such as the high Q-factors, and simultaneously remove the obstacles hindering their use, namely, the WG-mode double degeneracy. In this work, for the first time to our knowledge, it is shown (through 2D numerical simulations based on the presented here rigorous integral-equation method) that optimally tuned PMs also demonstrate significant enhancement of the sensitivity to the changes in their environment. Furthermore, we demonstrate that arranging such PMs into symmetrical "super-molecules" results in even further enhancement of their Q-factors and sensitivity, paving a way for their use as sensitive nano-scale biochemical sensors on a chip. Finally, we show that such PMs also enable detection of individual nanoparticles with the sizes below the diffraction limit.

**Problem geometry and methodology**

Photonic molecules composed of several evanescently-coupled microcavities supporting high-Q WG modes play an important role in photonics research as they can potentially find use for various applications ranging from optical power transfer via coupled resonator optical waveguides to low-threshold PM microlasers. A photonic molecule composed of $L$ side-coupled microdisk resonators is considered as illustrated in Fig. 1. The dielectric (or semiconductor) microdisks of radii $a_l$ and permittivities $\varepsilon_l$ ($l=1...L$) are located in a host medium with permittivity $\varepsilon_e$.

For thin disks, the three-dimensional problem of finding the eigenfrequencies of the photonic molecule can be replaced with an equivalent 2-D formulation by using the effective-index method to account for the vertical field confinement (the disk permittivities should then be replaced with their corresponding effective values)[18]. In the rest of this paper, we will thus restrict ourselves to simplified 2-D calculations. The effective refractive index of the disks used in the following sections ($n_{eff}$ = 2.63) may correspond to e.g. air-clad 165-nm thick silicon disks ($n$ = 3.52) located on a silica substrate ($n$ = 1.444) or 200-nm thick InGaAsP disks ($n$ = 3.37) suspended in the air at 1550 nm. To account for the material losses in the cavity, we also assume a small positive value of the imaginary part of the cavity effective dielectric constant ($\varepsilon_{eff}$ = 6.9169+$i$10$^{-4}$). Applying the Green's formula to the fields ($U_l$) and the Green's functions in the regions inside ($G_l^c$) and outside ($G_l^e$) all the microcavities and taking into account the transparency conditions at the cavity boundaries, we can formulate the problem in terms of the 2-D Muller boundary integral equations (MBIEs)[18]:

$$U_p(\mathbf{r}) = U^{inc}(\mathbf{r}) +$$
$$\sum_{l=1}^{L} \int_{S_l} \left[ U_l(\mathbf{r}') \frac{\partial}{\partial n'} \left( G_l^c(\mathbf{r},\mathbf{r}') - G_l^e(\mathbf{r},\mathbf{r}') \right) - V_l(\mathbf{r}') \left( G_l^c(\mathbf{r},\mathbf{r}') - \frac{\alpha_e}{\alpha_l} G_l^e(\mathbf{r},\mathbf{r}') \right) \right] ds_l'$$

$$\frac{\alpha_e + \alpha_p}{2\alpha_p} V_p(\mathbf{r}) = V^{inc}(\mathbf{r}) +$$
$$\sum_{l=1}^{L} \int_{S_l} \left[ U_l(\mathbf{r}') \frac{\partial^2}{\partial n \partial n'} \left( G_l^c(\mathbf{r},\mathbf{r}') - G_l^e(\mathbf{r},\mathbf{r}') \right) - V_l(\mathbf{r}') \frac{\partial}{\partial n} \left( G_l^c(\mathbf{r},\mathbf{r}') - \frac{\alpha_e}{\alpha_l} G_l^e(\mathbf{r},\mathbf{r}') \right) \right] ds_l'$$

(1)

where $p = 1...L$ ($L$ being a total number of cavities), and

$$\begin{array}{l} U_l^{TE}(\mathbf{r}) = H_z^l(\mathbf{r}),\ V_l^{TE}(\mathbf{r}) = \frac{\partial H_z^l(\mathbf{r})}{\partial n},\ \alpha_l = \varepsilon_l,\ \alpha_e = \varepsilon_e \\ U_l^{TM}(\mathbf{r}) = E_z^l(\mathbf{r}),\ V_l^{TM}(\mathbf{r}) = \frac{\partial E_z^l(\mathbf{r})}{\partial n},\ \alpha_l = \mu_l,\ \alpha_e = \mu_e \end{array},\ l = 1...L \qquad (2)$$

Note that the analytical procedure developed here for the case of $L$ electromagnetically coupled microcavities is a generalization of the solution for a single isolated microcavity[18]. Equations (1) are the Fredholm second-kind integral equations with either smooth or integrable kernels and are free of spurious solutions. Thus, their discretization yields well-conditioned matrices[18].

Note that formulation (1) is valid for microdisks of arbitrary smooth cross-sections. As the next step of the solution procedure, we shall obtain a discretized form of the MBIEs for a specific (and important for a number of practical applications) case of $L$ coupled circular microdisks of different radii $a_l$. The expressions for the Green's functions for the case of the source point Q located inside and the observation point P located either inside or outside of the $p$-th microcavity of radius $a_p$ and permittivity $\varepsilon_p$ immersed into a host medium with permittivity $\varepsilon_e$ can be written in the coordinate system associated with the $p$-th cavity as follows (see Fig. 2a):

$$G_p^c(\mathbf{r},\mathbf{r}') = \frac{i}{4} \sum_{(m)} J_m(k_p r_p(\text{P})) H_m^{(1)}(k_p r_p(\text{Q})) \exp(-im\theta_p(\text{Q})) \exp(im\theta_p(\text{P})),\ r_p(\text{P}) \leq r_p(\text{Q})$$

(3)

$$G_p^e(\mathbf{r},\mathbf{r}') = \frac{i}{4} \sum_{(m)} J_m(k_e r_p(\text{Q})) H_m^{(1)}(k_e r_p(\text{P})) \exp(-im\theta_p(\text{Q})) \exp(im\theta_p(\text{P})),\ r_p(\text{P}) \geq r_p(\text{Q})$$

However, if the source point Q is located inside the $l$-th cavity (of radius $a_l$ and permittivity $\varepsilon_l$) and the observation point P is located either inside the $p$-th cavity or in the outer region, the Green's functions take the following form in the local coordinate system associated with the $l$-th cavity (see Fig. 2b):

$$G_l^c(\mathbf{r},\mathbf{r}') = \frac{i}{4}\sum_{(m)} J_m(k_l r_l(Q))H_m^{(1)}(k_l r_l(P))\exp(-im\theta_l(Q))\exp(im\theta_l(P)), \quad r_l(P) \geq r_l(Q)$$

$$G_l^e(\mathbf{r},\mathbf{r}') = \frac{i}{4}\sum_{(m)} J_m(k_e r_l(Q))H_m^{(1)}(k_e r_l(P))\exp(-im\theta_l(Q))\exp(im\theta_l(P)), \quad r_l(P) \geq r_l(Q)$$

(4)

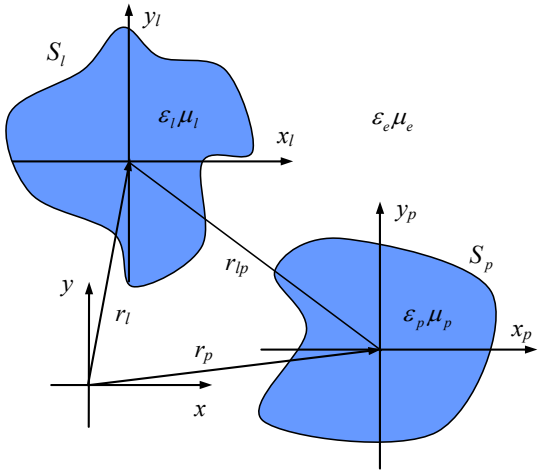

Fig. 1. Schematic of a coupled-microdisk cluster immersed into a homogeneous dielectric medium.

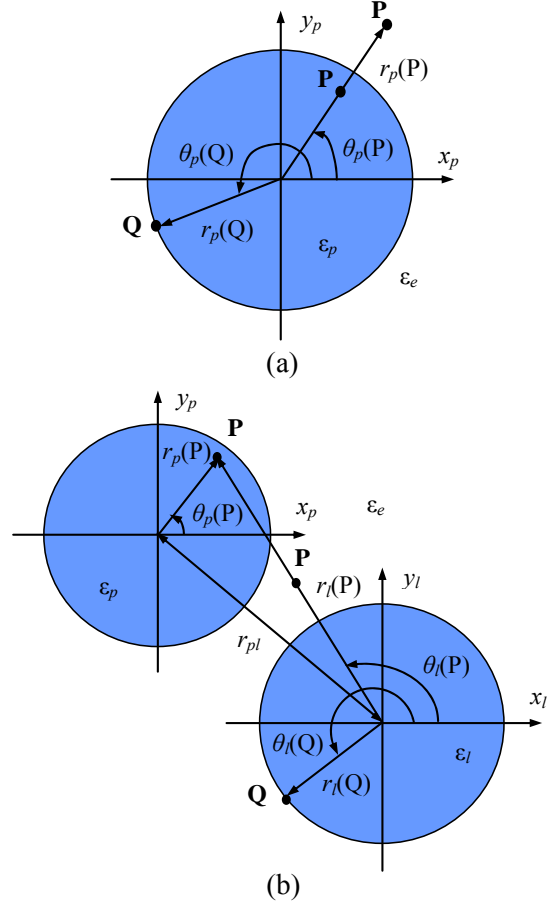

Fig. 2. Two possible locations of the source (Q) and the observation (P) points in the expressions for the Green's functions: (a) both points belong to the same cavity and (b) the points belong to different cavities. The global and local coordinate systems used in the analysis are also shown.

By using the following Graf's formulas for the Hankel functions[19,20], expressions (4) can be rewritten in a more convenient form, where the coordinates of the point P are given in the *p*-th coordinate system:

$$H_m^{(1)}(kr_l(P))\exp(im\theta_l(P)) =$$
$$\sum_{(n)} H_{n-m}^{(1)}(kr_{pl})J_n(kr_p(P))\exp(i(m-n)\theta_p^l)\exp(in\theta_p(P)), \quad r_p(P) < r_{pl}$$

$$H_m^{(1)}(kr_l(P))\exp(im\theta_l(P)) =$$
$$\sum_{(n)} J_{n-m}(kr_{pl})H_n^{(1)}(kr_p(P))\exp(i(m-n)\theta_p^l)\exp(in\theta_p(P)), \quad r_p(P) > r_{pl}$$

(5)

Now we substitute the above expressions into the MBIEs (1), and expand the unknown functions in terms of the Fourier series with angular exponents as global basis functions. Testing against the same set of global functions yields the final matrix equation that needs to be solved to determine the unknown Fourier coefficients $u_m^p$ and $v_m^p$:

$$a_m^p u_m^p + b_m^p v_m^p + \sum_{l \neq p} \left\{ \sum_{(n)} u_n^l A_{mn} + \sum_{(n)} v_n^l B_{mn} \right\} = 0$$

$$c_m^p u_m^p + d_m^p v_m^p + \sum_{l \neq p} \left\{ \sum_{(n)} u_n^l C_{mn} + \sum_{(n)} v_n^l D_{mn} \right\} = 0$$

(6)

where

$$a_m^p = \sqrt{\varepsilon_p} J_m(k_p a_p) H_m^{(1)'}(k_p a_p)$$
$$- \sqrt{\varepsilon_e} J_m'(k_e a_p) H_m^{(1)}(k_e a_p) + \frac{4}{i\pi k a_p}$$

$$b_m^p = J_m(k_p a_p) H_m^{(1)}(k_p a_p) - \frac{\alpha_e}{\alpha_p} J_m(k_e a_p) H_m^{(1)}(k_e a_p)$$

$$c_m^p = \varepsilon_e J_m'(k_e a_p) H_m^{(1)'}(k_e a_p) - \varepsilon_p J_m'(k_p a_p) H_m^{(1)'}(k_p a_p)$$

$$d_m^p = \frac{\alpha_e}{\alpha_p} \sqrt{\varepsilon_e} J_m(k_e a_p) H_m^{(1)'}(k_e a_p)$$
$$- \sqrt{\varepsilon_p} J_m'(k_p a_p) H_m^{(1)}(k_p a_p) + \frac{2(\alpha_p + \alpha_e)}{i\pi \alpha_p k a_p}$$

and

$$A_{mn} = \left( \sqrt{\varepsilon_l} J_n'(k_l a_l) J_m(k_l a_p) H_{m-n}^{(1)}(k_l r_{pl}) \right.$$
$$\left. - \sqrt{\varepsilon_e} J_n'(k_e a_l) J_m(k_e a_p) H_{m-n}^{(1)}(k_e r_{pl}) \right) e^{i(n-m)\theta_p^l}$$

$$B_{mn} = \left( J_n(k_l a_l) J_m(k_l a_p) H_{m-n}^{(1)}(k_l r_{pl}) \right.$$
$$\left. - \frac{\alpha_e}{\alpha_l} J_n(k_e a_l) J_m(k_e a_p) H_{m-n}^{(1)}(k_e r_{pl}) \right) e^{i(n-m)\theta_p^l}$$

$$C_{mn} = \left(\varepsilon_e J'_n(k_e a_l) J'_m(k_e a_p) H^{(1)}_{m-n}(k_e r_{pl})\right.$$
$$\left. - \varepsilon_l J'_n(k_l a_l) J'_m(k_l a_p) H^{(1)}_{m-n}(k_l r_{pl})\right) e^{i(n-m)\theta^l_p}$$

$$D_{mn} = \left(\frac{\alpha_e}{\alpha_l}\sqrt{\varepsilon_e} J_n(k_e a_l) J'_m(k_e a_p) H^{(1)}_{m-n}(k_e r_{pl})\right.$$
$$\left. - \sqrt{\varepsilon_l} J_n(k_l a_l) J'_m(k_l a_p) H^{(1)}_{m-n}(k_l r_{pl})\right) e^{i(n-m)\theta^l_p}$$

Coefficients $a_m$-$d_m$ of matrix (6) correspond to the matrix coefficients of a problem for an isolated $p$-th cavity, while coefficients $A_{mn}$-$D_{mn}$ describe the optical coupling between the $p$-th and the $l$-th cavities. It should be noted here that if several microcavities are arranged in a PM with the inter-cavity spacing comparable to the cavities sizes and the optical wavelength, these coefficients play a very important role in the numerical solution, as mutual electromagnetic interactions inside the structure can drastically affect its optical properties. Such PMs support collective super-modes, whose frequencies and Q-factors may vary significantly from those of a single microcavity. The algorithm presented here yields analytical expressions for all the matrix coefficients and thus provides superior accuracy of numerical solutions.

Homogeneous equation (6) has nonzero solutions only at frequencies where the equation matrix becomes singular. Complex eigenfrequencies of the optical modes supported by a set of coupled resonators are found by searching for the roots of the matrix determinant in the complex frequency plane. Once the eigenfrequencies are found, the modal fields and emission characteristics can be calculated by solving the homogeneous matrix equation (6) at the eigenfrequencies. In the frame of the 2−D model, the numerical results are exact up to the truncation error of angular Fourier series, and up to the accuracy of the root searching in the complex plane (both operations have been performed with machine precision).

## Q-factor and FSR increase in the optimally tuned PM

As it has been mentioned in the introduction, microcavity-based sensors either detect the presence of a gas or an analyte in the cavity cladding by measuring the resonant mode wavelength shift (mass sensing) or measure the enhanced fluorescence from the cladding material at the resonant wavelength of the microcavity (fluorescence sensing). In both types of applications, high value of the Q-factor of the cavity mode used for detection is crucial for efficient and robust detection, as the resonance linewidth and the fluorescence enhancement effect are directly related to this value.

Whispering-gallery modes supported by microdisks are classified as WG$_{mnl}$ modes, $m$ being the azimuthal number, $n$ the radial number, and $l$ the vertical number. In thin disks considered in this paper, $l = 1$ and will be omitted in the following discussion. Furthermore, only TE-polarized (having electric field mostly in the cavity plane) modes will be considered as in thin disks they have much larger effective refractive index values and are dominant[11]. The field in the 2D setting is thus determined by a scalar principal vertical (out of plane) magnetic field component $H_z$. It is known that theoretical values of the Q-factors of the circular microdisk WG modes grow exponentially with increased disk radius (increased modes azimuthal number)[11]. This dependence is clearly observed in Fig. 3, where the Q-factors of the WG modes in circular microdisks with increasing diameters are plotted. All the modes presented in Fig. 3 have resonant wavelengths around λ = 1.55 μm. However, such exponential Q-factor grows is not observed in real-life microdisks, as high-Q values of higher-azimuthal-order modes are spoiled by the cavity sidewall surface roughness[12,21].

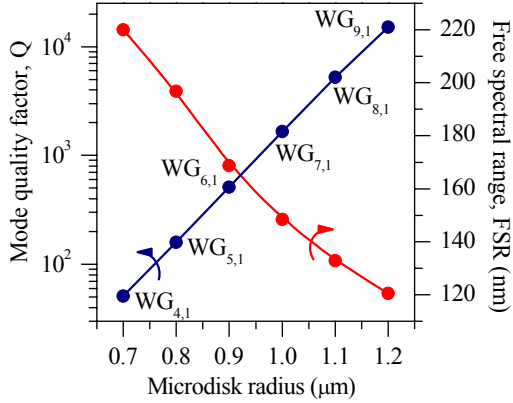

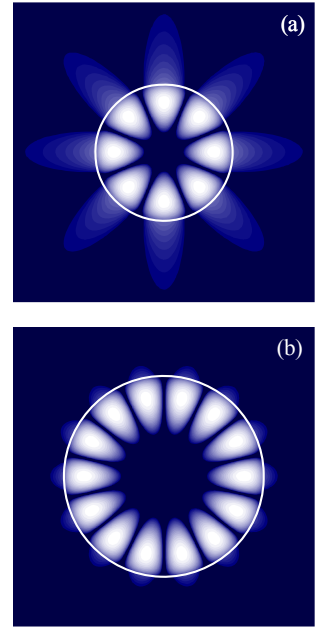

Fig. 3. Values of the quality factors and FSRs of the WG-modes in a 2-D circular microdisk resonator ($n_{eff}$ = 2.63, $n_e$ = 1.0) with resonant wavelengths around 1.55 μm as a function of the disk radius. Note that all the modes are double-degenerate due to the disk symmetry.

Fig. 4. Magnetic near-field distributions of the WG-modes in the vertical cross sections of two resonators: (a) $WG_{4,1}$ mode in a 1.34 μm diameter disk and (b) $WG_{7,1}$ mode in a 2 μm diameter disk.

Another important cavity characteristic, the free spectral range (FSR) or mode spectral spacing is also plotted in Fig. 3. Large FSR value is especially important for the fluorescence sensing as solvated fluorescent molecules usually have large emission bandwidths. Note that the cavity FSR should be much greater than the emission spectral width in order to yield the net fluorescence enhancement[9,22,23]. Unlike the Q-factors, the FSR values decrease with the increased cavity size (Fig. 3). It should also be noted that if the cavity surface roughness or coupling to a bus waveguide lifts the WG-mode degeneracy, the FSR values become much smaller.

Finally, it should be taken into account that the WG-mode interaction with the analyte occurs only through the evanescent portion of the mode field. Optical field of the high-Q WG modes is very efficiently confined in the microdisk and therefore these modes are not very sensitive to the small changes in the microcavity environment. Near-field distributions (portraits of $|H_z(x,y)|$) of the $WG_{4,1}$ mode in a smaller 1.34-μm diameter microdisk, and of the $WG_{7,1}$ mode in a larger 2 μm-diameter cavity are presented in Figs. 4 a and b, respectively. It can be clearly seen that the $WG_{4,1}$ mode field extends much farther into the surrounding medium in the cavity plane. This presence of the region of relatively high field intensity outside the cavity makes the cavity characteristics sensitive to the presence of the analyte in the cladding. At the same time, the poor mode confinement results in the very low $WG_{4,1}$ mode Q-factor value, Q=50.8 (for comparison, the Q-factor of the $WG_{7,1}$ mode is 1660).

To overcome this contradiction, this paper suggests arranging low-Q high-FSR nanocavities into pre-designed symmetrical configurations supporting symmetry-enhanced photonic molecule super-modes with high Q-factors. For example, a square PM configuration has already been shown to offer significantly (over 20 times) enhanced Q-factor of the $WG_{6,1}$ molecule mode having odd field symmetry along the diagonals and even symmetry along the x- and y- axes (OE-mode)[14]. As it is demonstrated in Fig. 5, this Q-factor boost can be achieved by optimally tuning the inter-cavity coupling distance. Fig. 5 a,b shows the dependence of the resonant wavelength shifts and Q-factor enhancements of three PM supermodes with four, six, and eight azimuthal field variations on the distance from the cavities centers to the PM center. Clearly, the lower the cavity azimuthal WG-mode number, the more sensitive it is to the presence of other cavities forming the photonic molecule. This higher sensitivity results in even more dramatic Q-factor enhancement (over 50 times) of the $WG_{4,1}$-supermode of OE symmetry ($Q^{OE} = 2768$ for $d_m = 1.101$ μm). A near-field intensity portrait of the $WG_{4,1}$ OE -supermode is shown in Fig. 5 c. It should be also noted that for this optimal PM configuration all the other closely located PM supermodes of different symmetries have Q-factor values many times lower, so that the PM free spectral range is equal to that of a single cavity.

Next, in Fig 6a we observe that arranging four optimally-tuned PMs (with $d_m = 1.101$ μm) into a square configuration – we call it a photonic super-molecule (PSM) – and again adjusting the coupling distance $d_s$ yields even further enhancement of the Q-factor of the PSM OE-supermode ($Q^{OE} = 6656$ for $d_s = 2.216$ μm). The optimal configuration of the sixteen-cavity photonic super-molecule together with the PSM OE-supermode intensity portrait is shown in Fig. 6 b. Note that the PSM is not a periodic microdisk array but rather a fractal-like structure: the distances between four PMs ($d_s$) and between four microdisks composing each PM ($d_m$) have been tuned separately, and the inter-cavity separations are not identical for all the neighboring cavities. Once again, the FSR of the optimally-tuned PSM is equal to its single-cavity counterpart.

## Sensitivity enhancement at PM resonances

As PM modes are collective multi-cavity resonances, we expect their resonant frequencies to be more sensitive to changes in their environment (e.g., variations of the refractive index in the inter-cavity region) than the corresponding characteristics of a larger single cavity with a comparable value of the Q-factor. In general, a microcavity configuration that provides a better overlap of the modal field with the analyte without degrading the mode Q factor should offer enhanced sensitivity[6].

Now, we compare the shifts of the resonant frequencies of a single microdisk WG mode and of the symmetry-enhanced OE supermodes of smaller-radius disks arranged into optimally-tuned configurations caused by the change of the refractive index of the host medium. As a test single-cavity structure, we chose a 2 μm-diameter microdisk resonator supporting a $WG_{7,1}$ mode with the Q-factor value (Q = 1660) of the same order as the PM OE-supermode Q-factor. $WG_{8,1}$ mode in a larger-radius microdisk will have a higher Q-factor value (see Fig. 3) but will be even less sensitive to the changes in the refractive index outside the resonator. As illustrated in Fig. 7a, the cavity mode wavelengths linearly shift to higher values as the refractive index value outside the resonator (i.e., gas or solution concentration) is increased. Clearly, the refractive index sensitivity (determined by the slope of the lines in Fig. 7a) is enhanced for the PM-based sensor and even further enhanced for the PSM-based one. Also note (Fig. 7b) that although all the modes Q-factors degrade slowly due to lowered index contrast between the microcavity and the cladding material, the PM and PSM Q-factors remain higher than the single-cavity mode Q-factor. It can also be expected that the increased overlap of the evanescent field of the high-Q PM and SPM modes with the surrounding medium would yield enhanced performance of biosensors based on the PMs immersed in a fluorophore solution.

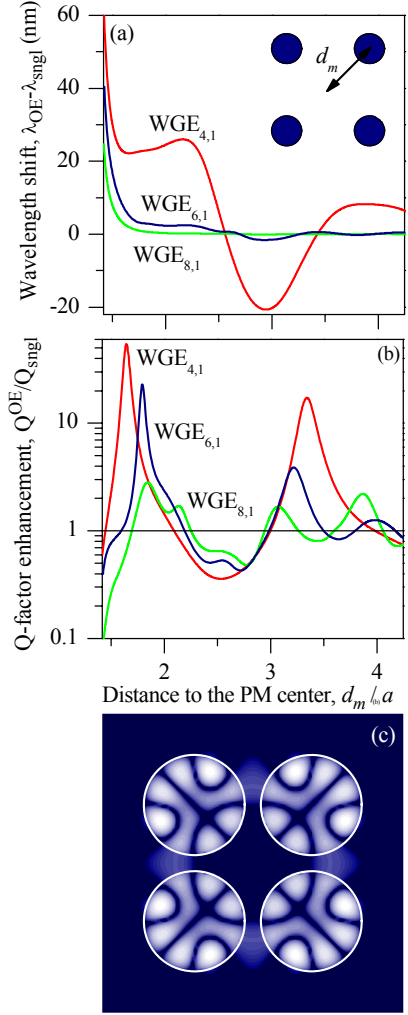

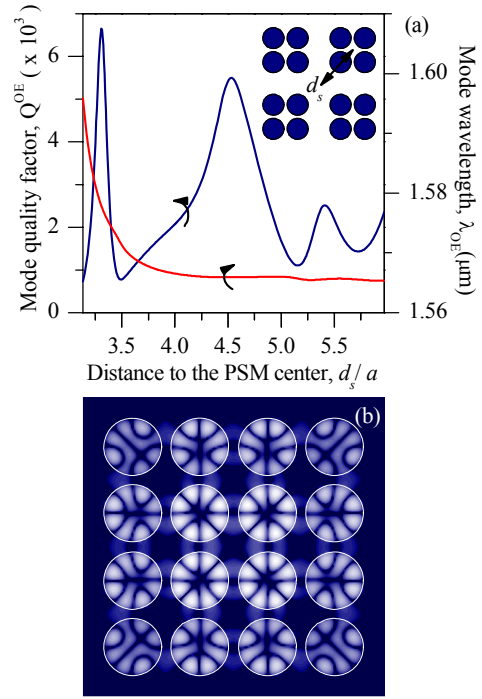

Fig. 5. (a) Wavelength shifts and (b) Q-factor enhancements relative to corresponding single-cavity values of three OE-supermodes of different azimuthal orders in a square PM vs distance $d_m$, and (c) near-field intensity distribution of the $WG_{4,1}$ OE-supermode for $d_m$ = 1.101 μm. The disk radii are $a$ = 0.67 μm, 0.9 μm, and 1.1 μm, respectively. The inset shows the PM geometry.

Fig. 6. (a) Wavelength shift and Q-factor change of the $WG_{4,1}$ OE-supermode in a square PSM vs distance $d_s$ and (b) near-field intensity distribution of the $WG_{4,1}$ OE-supermode in the optimally-tuned PSM configuration with $d_s$ = 2.216 μm.

Next, to calibrate the PM-based sensor for the detection of various solvents in the cladding of the structure, we consider a square PM consisting of the water-clad ($n = 1.333$) 200-nm thick, 1.3 µm-diameter silicon disks ($n = 3.52$) located on a silica substrate ($n = 1.444$). The effective refractive index of the microcavities is $n_{eff} = 2.813$. The optimally-tuned PM configuration yielding the symmetry-enhanced $WG_{4,1}$ OE-supermode corresponds to $d_s = 1.078$ µm. As illustrated in Fig. 8, the resonant wavelength of the OE-supermode of a photonic molecule systematically shifts to higher values with the increase of the refractive index of the solvent. The points on the plot correspond to the refractive indices of (a) pure water[3], (b) 0.5% glucose solution[24], (c) 10% glucose solution, (d) EtOH[3], (e) 1-propanol[25]. For example, changing of the cladding refractive index from that of water ($n = 1.3330$) to alcohol ($n = 1.3614$) and to 1-propanol ($n = 1.3845$) yields significant resonant wavelength upshifts of 4.28 nm and 7.88 nm, respectively (Fig. 8). Linear regression analysis yields a value of the refractive index sensitivity of 153 nm RIU$^{-1}$. This theoretically predicted value of the refractive index sensitivity is of the same order as the experimentally reported typical sensitivity values of surface plasmon biosensors based on individual chemically synthesized Ag nanoparticles (160-235 nm RIU$^{-1}$)[25] and as the values reported for nanolithography-fabricated Ag nanoparticle arrays (191 nm RIU$^{-1}$)[26]. Note, however, that the Q-factors of the surface plasmon resonances on noble-metal nanoparticles are much lower than corresponding values of the optical microcavity WG-modes. For comparison, the measured refractive index sensitivity of a sensor based on a single $Si_3N_4$ circular microdisk resonator[3] ($Q = 4900$) is only 22.89 nm RIU$^{-1}$.

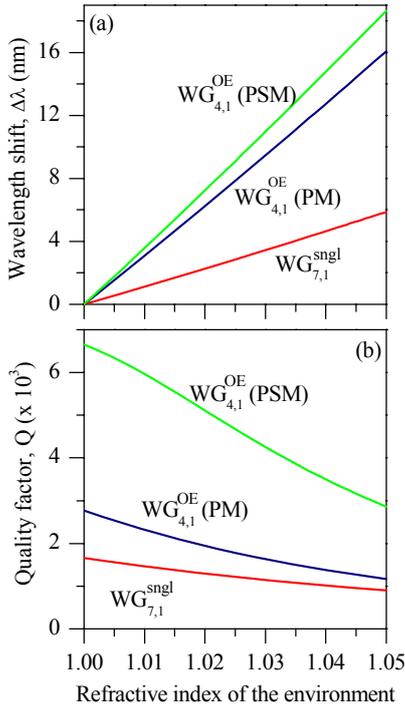

Fig. 7. Comparison of the wavelength shifts $\Delta\lambda = \lambda(n_e)-\lambda_0$ (a) and Q-factors change (b) as a function of the refractive index of the environment for a single 2 µm diameter microdisk operating on a $WG_{7,1}$ mode and for the PM and PSM operating on the symmetry-enhanced $WG_{4,1}$ OE-supermodes.

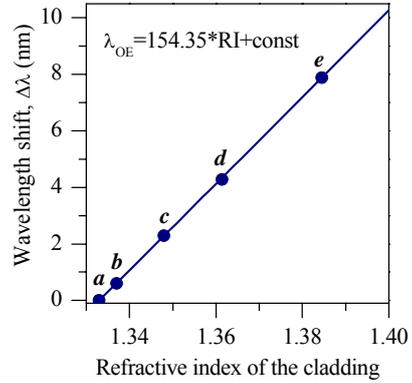

Fig. 8. Calculated shifts of the PM $WG_{4,1}$ OE-supermode resonant wavelength as a function of the refractive index of the sensor cladding: (a) pure water, (b) 0.5% glucose, (c) 10% glucose, (d) EtOH, (e) 1-propanol.

## Single-particle detection

Finally, we explore a capability of the PM-based sensors to detect individual sub-wavelength sized nanoparticles (note that conventional optical microscopy is limited to a spatial resolution of about 200 mn, or a half of the wavelength of visible light) [27]. Similar to the cases of the mass or fluorescence sensing, the nanoparticle can interact with a microdisk or a PM mode field only through a small part of the mode expanding outside the resonator material as an evanescent field. This particle-WG mode coupling results in the frequency shift of the mode that can be measured to detect the presence of the nanoparticle. Fig. 9 presents wavelength shifts of the $WG_{4,1}$ OE-supermode of a square photonic molecule (Fig. 9a) and of the $WG_{7,1}$ mode of a single microcavity (Fig. 9b) caused by a presence of a single nanoparticle as a function of the nanoparticle radius. Two circular-shape dielectric nanoparticles are considered, with $n_{part}$ = 1.5 and $n_{part}$ = 2, and the microdisk and the PM parameters are the same as in Figs. 3-5. The nanoparticles are located either in between two neighboring cavities in the PM or at 1 nm distance from the outer microdisk rim (nanoparticle positions relative to the sensors are shown in the insets to Figs. 9a and b, respectively).

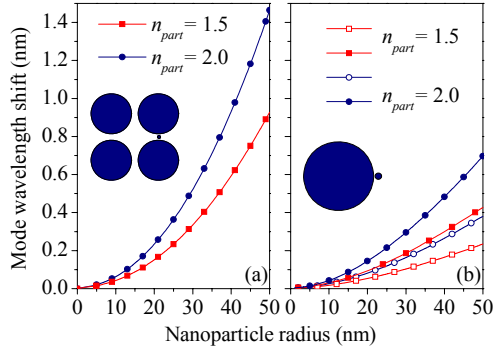

Fig. 9. Comparison of the wavelength shifts of the symmetry-enhanced PM $WG_{4,1}$ OE-supermode (a) and of the single-disk $WG_{7,1}$ mode (b) as a function of the radius of the detected nanoparticle. Degenerate microdisk mode splitting due to the presence of the nanoparticle is also observed in Fig. 9b. The insets show the nanoparticle positions relative to the microresonator structures.

Mode wavelengths shifts with the increased nanoparticle size are observed in both cases, though the PM supermode shows higher sensitivity to the presence of the nanoparticle. As it could be expected, higher-index nanoparticle causes more significant modes frequency shifts of both structures. However, because the presence of the nanoparticle close to the microdisk sidewall disturbs the circular symmetry of the structure, not only shift but also splitting of the microdisk WG-mode frequency can be observed in Fig. 9b. The wavelength spacing between the two split modes is about 0.2-0.3 nm, and this value drops rapidly with the increase of the nanoparticle-microcavity gap. Appearance of these two very closely located split modes with comparable Q-factors will complicate efficient detection of the nanoparticle. As the PM OE-mode is non-degenerate, no mode splitting occurs in Fig. 9a. Thus, optimally-tuned PM structures are expected to perform better that single microcavities in the sensor applications aimed at detection of single nanoparticles such as proteins, antibodies or viral particles as well as at monitoring processes of synthesis and aggregation[27].

Finally, accurate account for the evanescent-field interaction of a high-Q microcavity mode with a nano-particle (or a fiber tip) provided by the developed theoretical framework is important for a variety of applications, including near-field microscopy[28,29], lasing with a single nano-emitter such as an individual nano-crystal or a quantum dot[30], and enhancement of single-molecule fluorescence[31]. In order to correctly interpret the experimental results, the effect of the tip or the nano-emitter on the cavity optical mode frequency and Q-factor has to be carefully evaluated. Furthermore, in all these applications, when a tip or a nano-emitter is placed in the evanescent-field region of the microcavity it acts as a perturbing element that may yield degenerate cavity mode splitting (Fig. 9b). Thus, optimally-tuned quasi-single mode PM structures studied in this paper would offer improved performance over conventional symmetrical cavity geometries such as microdisks and microspheres.

## Fabrication tolerances and on-chip integration

To achieve the nanometer wavelength precision of high-index-contrast photonic molecule structures, nanometer geometry precision should be maintained. However, modern lithography techniques together with optimized etching processes enable fabrication of wavelength-scale optoelectronic components such as photonic wire waveguides, microring and microdisk resonators, and photonic-crystal circuits with very high and controllable precision[32]. For example, isolated microdisks[33] and large regular microdisk arrays[34] with diameters of and below 2 μm have been fabricated with high structural and optical quality, and their experimentally measured WG-mode frequencies and Q-factors have been reported to be remarkably close to the theoretical values predicted in the frame of simplified 2-D models[16,18,33,34]. Various techniques have also been developed for the reduction of the sidewall surface roughness (which is one of the principal sources of optical modes Q-factors degradation) such as, e.g., thermal oxidation of silicon-on-insulator photonic components[35]. Furthermore, lower-radial-order modes of smaller microdisks have been shown to exhibit better tolerance to the nanoscale surface roughness than high-radial-order high-Q WG modes in larger resonators[21].

Finally, by coupling optimally-tuned photonic molecules to bus waveguides, integrated semiconductor devices can be realized. In this case the $C_{4v}$ symmetry of the structure may be distorted, which can result in degrading the Q-factors of the PM super-modes. However, as it has been previously demonstrated for square-microdisk optical waveguide filters, this loss of $C_{4v}$ symmetry can be compensated and the high Q-factors maintained by slightly elongating the structure along the waveguide[36].

## Conclusions

Summarizing, we proposed use of coupled-microdisk photonic molecules that are tuned to support high-Q WG-supermodes as a possible biochemical sensor platform. Such PMs can be efficiently designed with the rigorous analytical method and the numerical algorithm presented in this paper. It is theoretically demonstrated that the resonant wavelengths of the supermodes of optimal PM configurations are much more sensitive to the changes of the refractive index of their environment than single-cavity WG-modes with comparable or even lower Q-factors. Furthermore, PM-based sensors show promise in detecting individual nanoparticles with the sizes below the diffraction limit, such as viral particles and proteins. We attribute the demonstrated sensitivity enhancement to the increased overlap of the evanescent fields of the PM modes with the detected biological or chemical material.

# References

1. R. W. Boyd and J. E. Heebner, "Sensitive disk resonator photonic biosensor," Appl. Opt. **40**(31) 5742-5747 (2001).
2. E. Krioukov, D. J. W. Klunder, A. Driessen, J. Greve, and C. Otto, "Integrated optical microcavities for enhanced evanescent-wave spectroscopy," Opt. Lett. **27**, 1504-1506 (2002).
3. E. Krioukov, D. J. W. Klunder, A. Driessen, J. Greve and C. Otto, "Sensor based on an integrated optical microcavity," Opt. Lett. **27**, 512–514 (2002).
4. C.-Y. Chao and L.J.Guo, "Biochemical sensors based on polymer microrings with sharp asymmetrical resonance," Appl. Phys. Lett. **83**(8) 1527-1529 (2003).
5. F. Vollmer, S. Arnold, D. Braun, I. Teraoka, A. Libchaber, "Multiplexed DNA quantification by spectroscopic shift of 2 microsphere cavities," Biophys. J. **85**, 1974–1979 (2003).
6. J. Scheuer, W. M. J. Green, G. A. DeRose, and A. Yariv, "InGaAsP annular Bragg lasers: theory, applications, and modal properties," IEEE J. Select. Topics Quantum Electron. **11**(2) 476-484 (2005).
7. H. Quan and Z. Guo, "Simulation of whispering-gallery-mode resonance shifts for optical miniature biosensors," J. Quantitative Spectroscopy Radiative Transfer **93**, 231–243 (2005).
8. W. Fang, D. B. Buchholz, R. C. Bailey, J. T. Hupp, R. P. H. Chang, and H. Cao, "Detection of chemical species using ultraviolet microdisk lasers," Appl. Phys. Lett. **85**(17) 3666-3668 (2004).
9. S. Blair and Y. Chen, "Resonant-enhanced evanescent-wave fluorescence biosensing with cylindrical optical cavities," Appl. Opt. **40**, 570–582 (2001).
10. H.-J. Moon, Y.-T. Chough, and K. An, "Cylindrical microcavity laser based on the evanescent-wave-coupled gain," Phys. Rev. Lett. **85**, 3161–3164 (2000).
11. R. E. Slusher, A. F. J. Levi, U. Mohideen, S. L. McCall, S. J. Pearton, and R. A. Logan, "Threshold characteristics of semiconductor microdisk lasers," Appl. Phys. Lett. **63**(10) 1310-1312 (1993).
12. M. Borselli, T. J. Johnson, and O. Painter, "Beyond the Rayleigh scattering limit in high-*Q* silicon microdisks: theory and experiment," Opt. Express **13**(5) 1515-1530 (2005).
13. M. Bayer, T. Gutbrod, J. P. Reithmaier, A. Forchel, T. L. Reinecke, and P. A. Knipp, "Optical modes in photonic molecules," Phys. Rev. Lett. **81**, 2582-2586 (1998).
14. S. V. Boriskina,"Theoretical prediction of a dramatic Q-factor enhancement and degeneracy removal of WG modes in symmetrical photonic molecules," Opt. Lett. **31**(3) 338-340 (2006).
15. E. I. Smotrova, A. I. Nosich, T. M. Benson, and P. Sewell, "Threshold reduction in a cyclic photonic molecule laser composed of identical microdisks with whispering-gallery modes," to appear in Opt. Lett. (2006).
16. A. Nakagawa, S. Ishii, and T. Baba, "Photonic molecule laser composed of GaInAsP microdisks," Appl. Phys. Lett. **86**, 041112 (2005).
17. A. L. Burin, H. Cao, G. C. Schatz and M. A. Ratner, "High-quality optical modes in low-dimensional arrays of nanoparticles: application to random lasers," J. Opt. Soc. Am. B **21**(1) 121-131 (2004).
18. S. V. Boriskina, P. Sewell, T. M. Benson, and A. I. Nosich, "Accurate simulation of 2D optical microcavities with uniquely solvable boundary integral equations and trigonometric-Galerkin discretization," J. Opt. Soc. Am. A **21**(3) 393-402 (2004).
19. M. Abramovitz and I. Stegun, Handbook of Mathematical Functions, Dover, New York (1970).

20. G. Tayeb and D. Maystre, "Rigorous theoretical study of finite-size two-dimensional photonic crystals doped by microcavities," J. Opt. Soc. Am. A **14**(12) 3323-3332 (1997).
21. S. V. Boriskina, T. M. Benson, P. Sewell, A. I. Nosich, "Spectral shift and Q-change of circular and square-shaped optical microcavity modes due to periodical sidewall surface roughness" J. Opt. Soc. Am. B **21**(10) 1792-1796 (2004).
22. H. Yokoyama and S. D. Brorson, "Rate equation analysis of microcavity lasers," J. Appl. Phys. **66**(10) 4801-4805 (1989).
23. M. D. Barnes, W. B. Whitten, S. Arnold, and J. M. Ramsey, "Enhanced fluorescent yields through cavity quantum electrodynamic effects in microdroplets," J. Opt. Soc. Am. B **11**, 1297-1304 (1994).
24. J. S. Maier, S. A. Walker, S. Fantini, M. A. Franceschini, and E. Gratton, "Possible correlation between blood glucose concentration and the reduced scattering coefficient of tissues in the near infrared," Opt. Lett. **19**, 2062-2064 (1994).
25. A. D. McFarland and R. P. Van Duyne, "Single silver nanoparticles as real-time optical sensors with zeptomole sensitivity," Nano Lett. **3**(8) 1057-1062 (2003).
26. M. D. Malinsky, K. L. Kelly, G. C. Schatz, and R. P. Van Duyne, "Chain length dependence and sensing capabilities of the localized surface plasmon resonance of silver nanoparticles chemically modified with alkanethiol self-assembled monolayers," J. Am. Chem. Soc. **123**(7) 1471-1482 (2001).
27. Z. Chen, A. Taflove and V. Backman, "Photonic nanojet enhancement of backscattering of light by nanoparticles: a potential novel visible-light ultramicroscopy technique," Opt. Express **12**(7) 1214-1220 (2004).
28. S. Gotziger, O. Benson, and V. Sandoghdar, "Influence of a sharp fiber tip on high-$Q$ modes of a microsphere resonator," Opt. Lett. **27**(2) 80-82 (2002).
29. A. Giusto, S. Savasta, and R. Saija, "Interaction of a microresonator with a nanoscatterer," IoP Publ. J. Phys.: Conf. Series **6**, 103-108 (2005).
30. M. Pelton and Y. Yamamoto, "Ultralow threshold laser using a single quantum dot and a microsphere cavity," Phys. Rev. A **59**, 2418 (1999).
31. M. Steiner, F. Schleifenbaum, C. Stupperich, A. V. Failla, A. Hartschuh, and A. J. Meixner, "Microcavity-controlled single-molecule fluorescence," ChemPhysChem **6**(10) 2190-2196 (2005).
32. R. W. Boyd, J. E. Heebner, N. N. Lepeshkin, Q.-H. Park, A. Schweinsberg, G. W. Wicks, A. S. Baca, J. E. Fajardo, R. R. Hancock, M. A. Lewis, R. M. Boysel, M. Quesada, R. Welty, A. R. Bleier, J. Treichler, and R. E. Slusher, "Nanofabrication of optical structures and devices for photonics and biophotonics," J. Mod. Opt. **50**(15-17) 2543-2550 (2003).
33. T. Baba, M. Fujita, A. Sakai, M. Kihara, and R. Watanabe, "Lasing characteristics of GaInAsP–InP strained quantum-well microdisk injection lasers with diameter of 2–10 μm," IEEE Photon. Technol. Lett. **9**(7) 878-880 (1997).
34. K. Petter, T. Kipp, Ch. Heyn, D. Heitmann, and C. Schuller, "Fabrication of large periodic arrays of AlGaAs microdisks by laser-interference lithography and selective etching," Appl. Phys. Lett. **81**(4) 592-594 (2002).
35. K. K. Lee, D. R. Lim, L. C. Kimerling, J. Shin, and F. Cerrina, "Fabrication of ultralow-loss Si/SiO2 waveguides by roughness reduction," Opt. Lett. **26**(23) 1888-1890 (2001).
36. M. Lohmeyer, "Mode expansion modeling of rectangular integrated optical microresonators," Opt. Quantum Electron. **34**(5) 541-557 (2002).